\documentclass[reprint,prl,twocolumn,showpacs,superscriptaddress,aps,longbibliography]{revtex4-2}

\usepackage[colorlinks=true,citecolor=blue]{hyperref} 
\usepackage{graphicx}
\usepackage{bm}
\usepackage{amsmath}
\usepackage{amssymb}

\DeclareGraphicsExtensions{.png,.jpg,.eps}
\usepackage{xcolor}

\newcommand{\beq}{\begin{eqnarray}}
\newcommand{\eeq}{\end{eqnarray}}
\begin{document}

\author{Guangze Chen}
\affiliation{Department of Applied Physics, Aalto University, 02150 Espoo, Finland}

\author{Fei Song}
\email{songf18@mails.tsinghua.edu.cn}
\affiliation{Institute for Advanced Study, Tsinghua University, Beijing, 100084, China}

\author{Jose L. Lado}
\email{jose.lado@aalto.fi}
\affiliation{Department of Applied Physics, Aalto University, 02150 Espoo, Finland}

\title{Topological spin excitations in non-Hermitian spin chains with a generalized kernel polynomial algorithm}

\begin{abstract}
Spectral functions of non-Hermitian Hamiltonians can reveal the existence of topologically non-trivial line gaps and the associated topological edge modes. However, the computation of spectral functions in a non-Hermitian many-body system remains an open challenge. Here, we put forward a numerical approach to compute spectral functions of a non-Hermitian many-body Hamiltonian based on the kernel polynomial method and the matrix-product state formalism. We show that the local spectral functions computed with our algorithm reveal topological spin excitations in a non-Hermitian spin model, faithfully reflecting the non-trivial line gap topology in a many-body model. We further show that the algorithm works in the presence of the non-Hermitian skin effect. Our method offers an efficient way to compute local spectral functions in non-Hermitian many-body systems with tensor-networks, allowing to characterize line gap topology in non-Hermitian quantum many-body models.
\end{abstract}

\date{\today}

\maketitle
Non-Hermitian (NH) Hamiltonians have risen as a rich framework to describe the post-selected quantum dynamics of open quantum systems\cite{breuer2002theory,Bergholtz2021RMP,Ashida2021}. NH Hamiltonians have been found to host a variety of phenomena lacking a counterpart in Hermitian systems,
including the non-Hermitian skin effect (NHSE)\cite{yao2018edge,lee2018anatomy,kunst2018biorthogonal,Xiao2019NHSE,Helbig2019NHSE,weidemann2020topological,Ghatak2019NHSE,PhysRevLett.116.133903,PhysRevB.97.121401}. 
In addition, NH Hamiltonians exhibit unique topological properties: a topologically non-trivial gap can either be a line gap or a point gap\cite{gong2018topological,kawabata2019symmetry,zhou2019periodic}. The line gap is a generalization of the spectral gap in Hermitian systems, and a topologically non-trivial line gap is associated with topological edge modes in open boundary conditions\cite{kawabata2019symmetry}. The point gap topology is unique for non-Hermitian systems and it encodes the origin of NHSE\cite{okuma2020topological,zhang2020correspondence,PhysRevLett.124.056802}. Recently, the generalization of NH topology as well as other interesting aspects of NH physics such as parity-time symmetry breaking\cite{bender1998real,bender2007making,el2018non} and exceptional points\cite{heiss2004exceptional,berry2004physics,Miri2019review,Ozdemir2019review} to NH many-body systems have attracted much interest\cite{PhysRevB.106.L121102,PhysRevResearch.1.033051,schafer2022symmetry,PhysRevB.105.165137,PhysRevResearch.4.033122}. 

Despite the intense interest in the study of NH many-body physics, efficient numerical tools
are lacking. 
In the case of interacting one-dimensional problems,
conventional tools to study Hermitian many-body systems based on
density matrix renormalization group (DMRG)\cite{PhysRevLett.69.2863}
cannot be directly applied to study NH many-body systems. 
In Hermitian many-body systems, local spectral functions can be
computed efficiently for large-size systems within the DMRG framework
using correction vector methods\cite{PhysRevB.60.335,PhysRevB.66.045114},
time-evolution\cite{PhysRevLett.93.076401,PhysRevLett.107.070601} or kernel polynomial
algorithms\cite{RevModPhys.78.275,PhysRevB.90.115124,PhysRevResearch.1.033009}. 
However, for NH many-body Hamiltonians, the kernel polynomial method (KPM) does not apply directly\cite{PhysRevE.94.063305}, and an efficient algorithm to compute local spectral functions in NH many-body systems is lacking.
The computation of local spectral functions of 
many-body excitations allow revealing many-body topological modes\cite{PhysRevResearch.1.033009},
and therefore would be of key interest to address topological
interacting NH many-body models, in particular with a nontrivial line gap. 

Here, we put forward a many-body non-Hermitian kernel polynomial method (NHKPM)\footnote{The source codes are available on \href{https://github.com/GUANGZECHEN/NHKPM.jl}{https://github.com/GUANGZECHEN/NHKPM.jl}} to compute local spectral functions in NH many-body systems. We implement a KPM-based algorithm to compute the local spectral functions, where matrix-product states (MPS) are used to represent the states. We show that this method reveals topological many-body spin excitations in a NH spin chain with non-trivial line gap topology.
We further show that this method provides faithful results in the presence of NHSE. 
Our results put forward a many-body algorithm to efficiently compute spectral functions in NH many-body systems, which in particular allows to characterize line gap topology in NH many-body systems.

We first summarize how KPM is used to compute local spectral functions in Hermitian spin systems. For a Hermitian $S=1/2$ Hamiltonian $H_0$, the local spectral density of $S_z=\pm1$ excitations is given by the local spin structure factor:
\beq \label{spin_structure_factor}
\begin{aligned}
\rho(E,l)=&\langle \text{GS} | S_l^{-}\delta(E+E_\text{GS}-H_0)S_l^{+} |\text{GS}\rangle\\
+&\langle \text{GS} | S_l^{+}\delta(E+E_\text{GS}-H_0)S_l^{-} |\text{GS}\rangle
\end{aligned}
\eeq
where $|\text{GS}\rangle$ and $E_\text{GS}$ are the ground state and its energy, and $S_l^{\pm}=S_l^x\pm iS_l^y$ where $l$ is the site index.
Such local spin structure factor reveals topological spin excitations\cite{PhysRevResearch.1.033009}, which can be probed with scanning tunneling microscopy \cite{Andreas2004,Otte2014,Andreas2015,PhysRevLett.119.227206,Shawu2019}. Since the eigenvalues of $H_0$ are bounded, we can perform a shifting and a rescaling on $H_0$ such that $E_\text{GS}=0$ and the eigenvalues lie in $[0,1)$. In such case, $\rho(E\in[0,1),l)$ is a bounded single-variable function for fixed $l$, which is a crucial property that allows it to be expanded in Chebyshev polynomials $T_n(E)=\cos(n\arccos{E})$:
\beq \label{eq2}
\rho(E,l)=\frac{1}{\pi\sqrt{1-E^2}}\left(\mu_0+2\sum_{n=1}^\infty\mu_nT_n(E)\right)
\eeq
where the coefficients satisfy $\mu_n=\int_{-1}^1\text{d}E\rho(E,l)T_n(E)=\langle \text{GS} | S_l^{-}T_n(H_0)S_l^{+} |\text{GS}\rangle
+\langle \text{GS} | S_l^{+}T_n(H_0)S_l^{-} |\text{GS}\rangle$. Using the recursion relation of Chebyshev polynomials: $T_{n+1}(x)=2xT_n(x)-T_{n-1}(x)$, the first term of $\mu_n$ can be computed as $\mu_n=\langle v|v_n\rangle$ where $|v\rangle=S_l^{+} |\text{GS}\rangle$ and 
\beq \label{eqvn}
\begin{aligned}
&|v_{n+1}\rangle=2H_0|v_n\rangle-|v_{n-1}\rangle\\
&|v_0\rangle=|v\rangle, |v_1\rangle=H_0|v\rangle,
\end{aligned}
\eeq
and similarly for the second term. The recursive structure of Eq. \eqref{eqvn} allows the efficient computation of $\mu_n$ and $\rho(E,l)$. Furthermore, its simple algebraic structure allows it to be implemented with MPS representation of states, where the ground state can be computed with DMRG. Finally, Eq. \eqref{eq2} is approximated with a truncation of the summation to $N^\text{th}$ order, where a Jackson kernel $g^N_n=\frac{(N-n+1)\cos\frac{n\pi}{N+1}+\sin\frac{n\pi}{N+1}\cot\frac{\pi}{N+1}}{N+1}$ is multiplied to the coefficients $\mu_n$ to damp Gibbs oscillations\cite{RevModPhys.78.275}.

We now generalize Eq. \eqref{spin_structure_factor} to an open spin system described with an effective NH spin Hamiltonian $H$. Unlike a Hermitian Hamiltonian, a NH Hamiltonian generally has a complex spectrum $\{E_n\in\mathbb{C}\}$. In addition, the left eigenvector $|\Psi_{\text{L},n}\rangle$ and the right eigenvector $|\Psi_{\text{R},n}\rangle$ corresponding to the same eigenvalue $E_n$ of $H$ are generally different and satisfy $\langle \Psi_{\text{L},m}|\Psi_{\text{R},n}\rangle=\delta_{mn}$. The eigen-decomposition of $H$ is: $H=\sum_nE_n|\Psi_{\text{R},n}\rangle\langle \Psi_{\text{L},n}|$. Thus, we define the local dynamical spin correlator as:
\beq \label{eq4}
\begin{aligned}
\mathcal{S}(\omega,l)=&\langle \text{GS}_\text{L} | S_l^{-}\delta^2(\omega+E_\text{GS}-H)S_l^{+} |\text{GS}_\text{R}\rangle\\
&+\langle \text{GS}_\text{L} | S_l^{+}\delta^2(\omega+E_\text{GS}-H)S_l^{-} |\text{GS}_\text{R}\rangle,
\end{aligned}
\eeq
where $\omega$ is a complex number, $|\text{GS}_\text{L,R}\rangle$ are left and right eigenvectors corresponding to the eigenvalue of smallest real part\cite{https://doi.org/10.48550/arxiv.2201.12653,PhysRevB.105.165137}, and $\delta^2(\omega-H)=\sum_n\delta(\Re(\omega-E_n))\delta(\Im(\omega-E_n))|\Psi_{\text{R},n}\rangle\langle\Psi_{\text{L},n}|$. Since $\omega$ is complex, $\mathcal{S}(\omega,l)$ is no longer a function of a single variable, and does not have the Chebyshev expansion Eq. \eqref{eq2} that allows it to be computed with KPM\footnote{There exists a multi-variable Chebyshev expansion for $f(\omega)$, yet the coefficients in such case cannot be computed efficiently as in Eq. \eqref{eqvn}\cite{RevModPhys.78.275}.}. 

We now present how an NHKPM algorithm\footnote{See Supplemental Material for the technical details of our algorithm and the analysis of the numerical complexity of our algorithm. Refs.\cite{https://doi.org/10.1112/plms/s1-29.1.519,PhysRevB.102.014203,SCHOLLWOCK201196,TEnss_2001,doi:10.1063/1.1899124,PhysRevB.101.235150,PhysRevB.105.205125} are included in the Supplemental Material.} allows to efficiently compute Eq. \eqref{eq4}, and more generally, generic spectral functions of the form:
\beq\label{eq5}
f(\omega)=\langle \psi_\text{L}|\delta^2(\omega-H)|\psi_\text{R}\rangle
\eeq
where $|\psi_\text{L}\rangle,|\psi_\text{R}\rangle$ are arbitrary states. The key observation is, although Eq. \eqref{eq5} cannot be computed with KPM directly, it can be related to the spectral functions of the Hermitrized form of $\omega-H$:
\beq
\mathcal{H}=\left(\begin{array}{cc}&\omega-H\\\omega^*-H^\dag&\end{array}\right).
\eeq
In particular, using $\partial_{\omega^*}(1/\omega)=\pi\delta(\Re(\omega))\delta(\Im(\omega))$\cite{FEINBERG1997579}, we can rewrite Eq. \eqref{eq5} as
\beq \label{eq7}
f(\omega)=\frac{1}{\pi}\partial_{\omega^*}G(E=0)
\eeq
where $G(E)=\langle\text{L}|(E-\mathcal{H})^{-1}|\text{R}\rangle$ is the Green's function of $\mathcal{H}$ with $|\text{L}\rangle=\left(\begin{array}{c}0\\|\psi_\text{L}\rangle\end{array}\right)$ and $|\text{R}\rangle=\left(\begin{array}{c}|\psi_\text{R}\rangle\\0\end{array}\right)$. Since $\mathcal{H}$ is Hermitian, its Green's function $G(E)$ has the Chebyshev expansion Eq. \eqref{eq2}, resulting in an expansion of $f(\omega)$ in Chebyshev polynomials of $\mathcal{H}$: 
\beq \label{eq8}
f(\omega)=\frac{2}{\pi^2}\sum_{n=1}^\infty(-1)^{n+1}\langle\text{L}|\partial_{\omega^*}T_{2n-1}(\mathcal{H})|\text{R}\rangle
\eeq
with the recursion relation $\partial_{\omega^*}T_{n+1}(\mathcal{H})=2\left(\begin{array}{cc}0&0\\1&0\end{array}\right)T_n(\mathcal{H})+2\mathcal{H}\partial_{\omega^*}T_{n}(\mathcal{H})-\partial_{\omega^*}T_{n-1}(\mathcal{H})$. Finally, the above recursion relation on $T_n(\mathcal{H})$ is transformed into an update of vectors, similar to Eq. \eqref{eqvn}, allowing $f(\omega)$ to be computed with KPM. We use
an MPS representation for the states $S_l^{\pm}|\text{GS}_\text{L,R}\rangle$ in the computation of Eq. \eqref{eq4}, which enables computation for large size systems. We use a Krylov-Schur algorithm\cite{doi:10.1137/S0895479800371529} to compute the states with the smallest real part\footnote{The fidelity of these states are verified in the Supplemental Material.}. 

\begin{figure}[t!]
\center
\includegraphics[width=\linewidth]{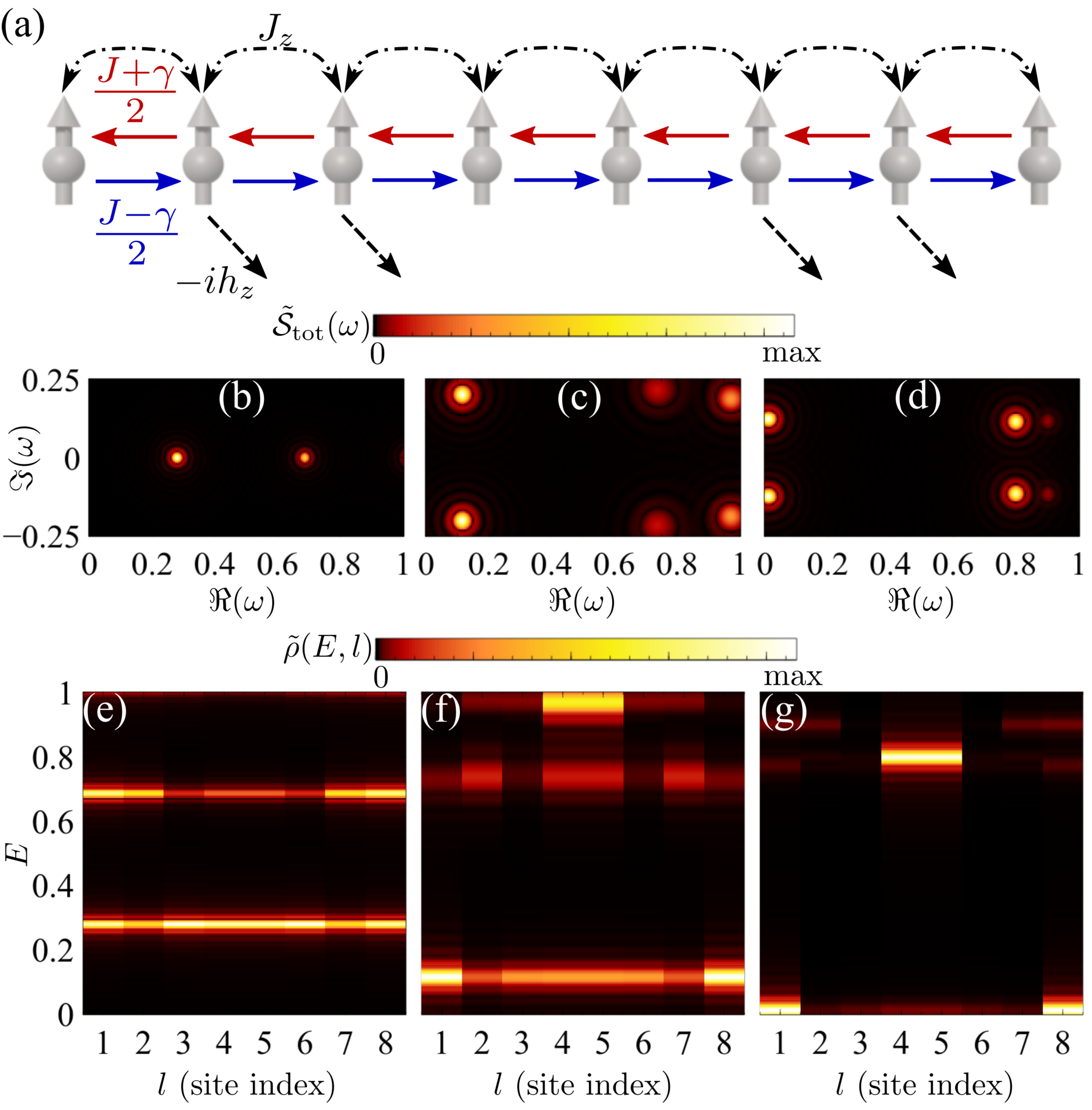}
\caption{(a) Sketch of the model Hamiltonian $H$ in Eq. \eqref{modeleq5}. (b-d) Total dynamical spin correlator $\tilde{\mathcal{S}}_\text{tot}(\omega)$ of $H$ with $L=8$, $\gamma=0$ and (b) $h_z=0$, (c) $h_z=J$, (d) $h_z=2J$. (e-g) Projected local dynamical structure factor $\tilde{\rho}(E,l)$ corresponding to cases (b-d), respectively. Topological spin excitations with almost zero real energy are revealed in (g).}
\label{fig1}
\end{figure}

We now use the NHKPM to identify the topological end modes emerging in a NH many-body system. We take 
the following NH $S=1/2$ 
chain with antiferromagnetic exchange
$J=1$ (Fig.\ref{fig1}(a)):
\beq\label{modeleq5}
\begin{aligned}
H&=\sum_{l=1}^{L-1}\left(\frac{J+\gamma}{2}S^+_lS^-_{l+1}+\frac{J-\gamma}{2}S^-_lS^+_{l+1}+J_zS^z_lS^z_{l+1}\right)\\&+\sum_{l=1}^Lih_l^zS_l^z
\end{aligned}
\eeq
where $h_l^z=-h_z$ if $l\mod4=2,3$ and $h_l^z=0$ otherwise. 
It is first illustrative to perform a Jordan-Wigner transformation to
Eq. \eqref{modeleq5}, leading to a NH interacting spinless fermion model:
\beq\label{model_fermion}
\begin{aligned}
\tilde{H}&=\sum_{l=1}^{L-1}\left(\frac{J+\gamma}{2}c^\dag_lc_{l+1}+\frac{J-\gamma}{2}c^\dag_{l+1}c_l\right.\\&\left.+J_z(c^\dag_lc_l-\frac{1}{2})(c^\dag_{l+1}c_{l+1}-\frac{1}{2})\right)+\sum_{l=1}^L ih_l^z(c^\dag_lc_l-\frac{1}{2}).
\end{aligned}
\eeq
When $J_z=0$, Eq. \eqref{model_fermion} becomes a non-interacting spinless fermion model, which for $\gamma=0$ and non-zero $h_z$ is known to give rise to topological end states whose real part of the energy is 0\cite{PhysRevB.100.161105}. 
The bulk topology of this model is characterized by a hidden Chern number which faithfully predicts the number of stable end states with purely imaginary energy\cite{PhysRevB.100.161105,PhysRevX.9.041015}. It has been shown that in the presence of a finite interaction $J_z$, a sufficiently large $h_z$ would still give rise to the topological end states\cite{PhysRevResearch.4.L012006}. In the following
we explore the non-Hermitian many-body topology of the model via computing dynamical excitations using the NHKPM algorithm, both in the absence and presence of NHSE. For concreteness we take $J_z=1/2J$, yet analogous results can be obtained for general values of $J_z$.

We first focus on a chain of length $L=8$ in which we can implement the NHKPM method exactly with exact diagonalization as a benchmark. We first compute the total dynamical spin correlator, defined as:
\beq \label{eq11}
\tilde{\mathcal{S}}_\text{tot}(\omega)=\left|\sum_{l=1}^L\mathcal{S}(\omega,l)\right|
\eeq
where $\mathcal{S}(\omega,l)$ is defined in Eq. \eqref{eq4} 
\footnote{We take the absolute value in Eq. \eqref{eq11} since $\mathcal{S}(\omega,l)$ is no longer real-valued in this case.}.
The total dynamical spin correlator reveals the energy of the states that have finite overlap with the ground state with one $S_z= \pm 1$ excitation. We see that as $h_z$ increases, the real part of the energy of the lowest excited states is shifted towards $0$, whereas the higher states are shifted away from $0$ (Figs.\ref{fig1}(b-d)). To see if the lowest excited states are topological end states, we compute the projected local dynamical structure factor, defined as
\beq \label{eq12}
\tilde{\rho}(E\in\mathbb{R},l)=\left|\int \mathcal{S}(E+iy,l)dy\right|.
\eeq
This quantity identifies the local spectral density of $S_z=\pm1$ excitations at a given real energy $E$. In particular, when $H$ is Hermitian, $\tilde{\rho}(E,l)$ is reduced to the local spin structure factor defined in Eq. \eqref{spin_structure_factor}. Figs.\ref{fig1}(e-g) shows $\tilde{\rho}(E,l)$ for different values of $h_z$, we see that for $h_z=0$ the lowest states show non-vanishing spectral density in the bulk. As $h_z$ increases to $h_z=J$ the spectral density in the bulk reduces, and eventually for sufficiently large $h_z=2J$ the states become localized at the ends, reflecting the non-trivial line topology in the bulk.
Due to the short chain length, finite-size effects prevent clearly observing topological end states for $h_z=J$ due to hybridization.

\begin{figure}[t!]
\center
\includegraphics[width=\linewidth]{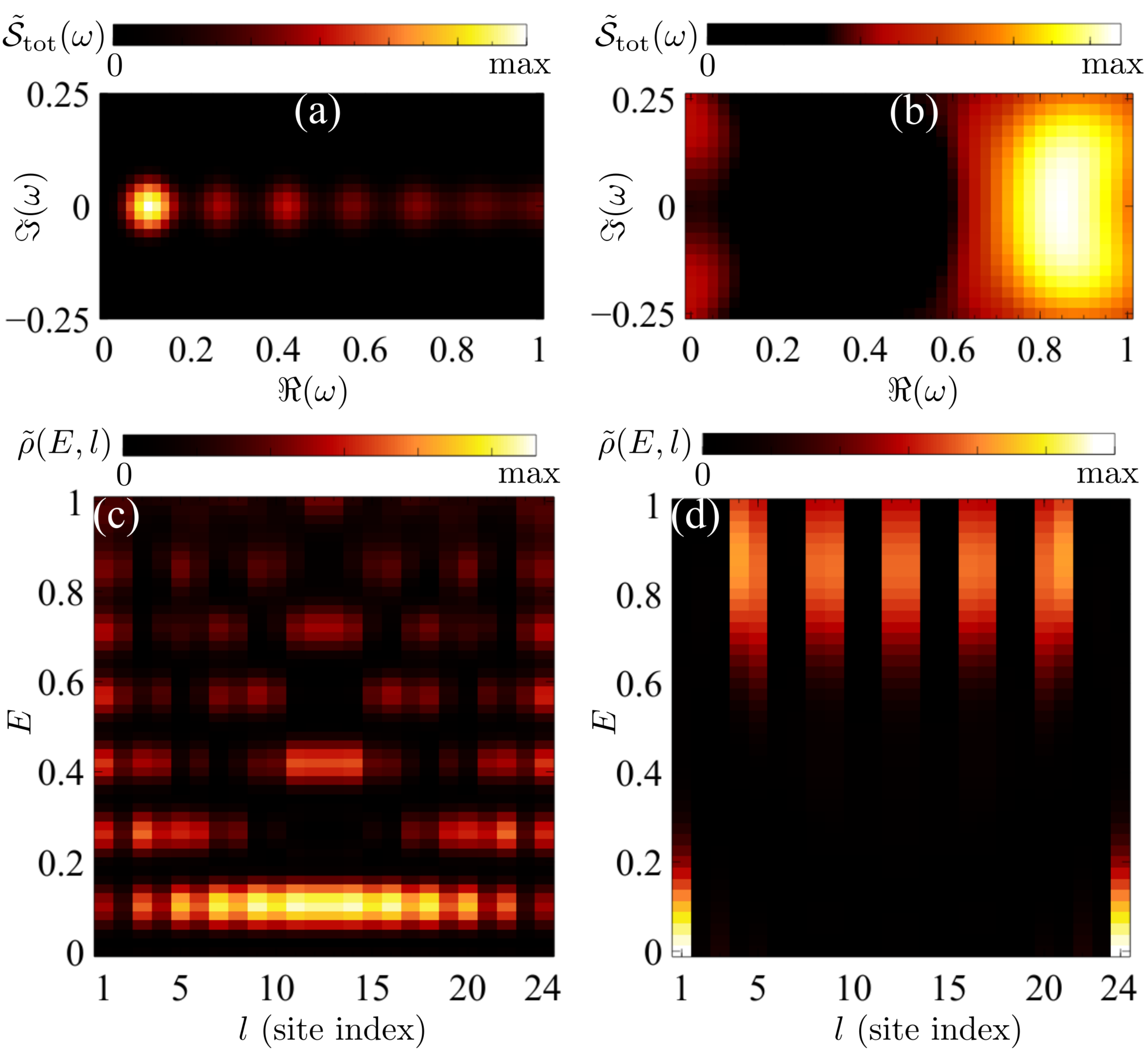}
\caption{(a,b) Total dynamical spin correlator $\tilde{\mathcal{S}}_\text{tot}(\omega)$ of $H$ defined in Eq.~\eqref{modeleq5} with $L=24$, $\gamma=0$ and (a) $h_z=0$ and (b) $h_z=J$. $\tilde{\mathcal{S}}_\text{tot}(\omega)$ reveals a many-body line gap in (b). (c,d) Projected local dynamical structure factor $\tilde{\rho}(E,l)$ corresponding to cases (a,b), respectively. $\tilde{\rho}(E,l)$ reveals topological many-body excitations in (d), indicating that the line gap in (b) is topological.}
\label{fig2}
\end{figure}

To approach the thermodynamic limit
and demonstrate the capabilities of the NHKPM
algorithm, we now move on
to consider a chain of length $L=24$.
As these systems are too large to be treated by exact diagonalization, we now use a full
tensor-network implementation with MPS in the NHKPM
algorithm. The total dynamical spin correlator $\tilde{\mathcal{S}}_\text{tot}(\omega)$ and the projected local dynamical structure factor $\tilde{\rho}(E,l)$ are computed for $h_z=0$ and $h_z=J$ in this case (Fig.\ref{fig2}). 
We see from Figs.\ref{fig2}(a,b) that for $h_z=J$ there are clearly 2 states close to 0 real energy, and are isolated from the higher states with a line gap, whereas for $h_z=0$ there is no such behavior. From Figs.\ref{fig2}(c,d) we can observe the topological end states for $h_z=J$, whereas for $h_z=0$ the lowest states are mainly distributed in the bulk. These show that the model Eq. \eqref{modeleq5} is not topological for $h_z=0$, and is topological for $h_z=J$ even in the presence of many-body interactions. 

\begin{figure}[t!]
\center
\includegraphics[width=\linewidth]{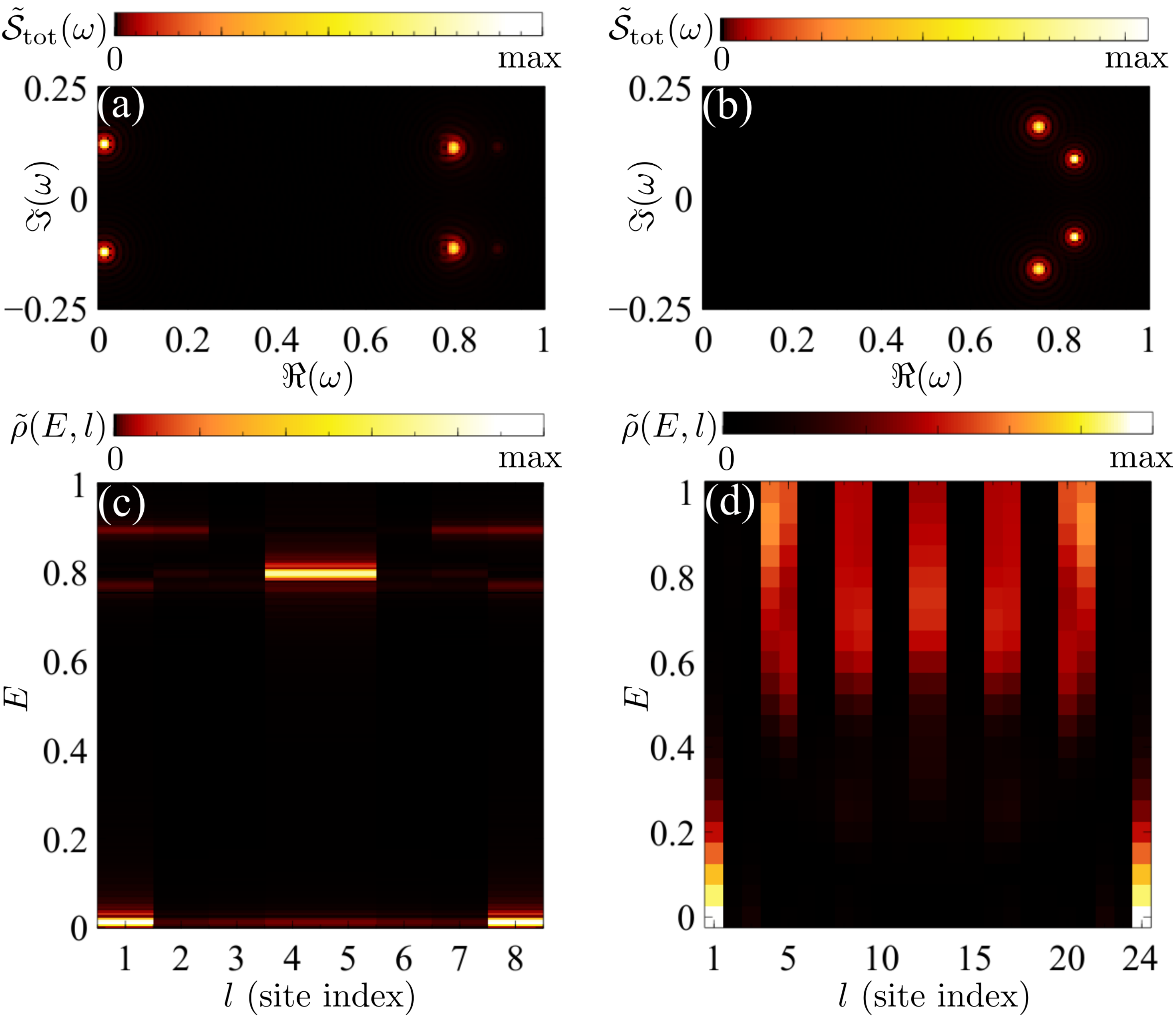}
\caption{(a,b) Total dynamical spin correlator $\tilde{\mathcal{S}}_\text{tot}(\omega)$ of $H$ defined in Eq.~\eqref{modeleq5} with $L=8$, $h_z=2J$ and $\gamma=0.1J$ under (a) open boundary condition and (b) periodic boundary condition. (c) Projected local dynamical structure factor $\tilde{\rho}(E,l)$ corresponding to (a), showing the persistence of topological end modes in the presence of NHSE. (d) $\tilde{\rho}(E,l)$ of $H$ with $L=24$, $h_z=J$ and $\gamma=0.1J$, showing the topological end modes in the presence of NHSE for a longer chain.}
\label{fig3}
\end{figure}

We now consider a finite $\gamma$ term, which is known to give rise to the NHSE\cite{yao2018edge,lee2018anatomy,kunst2018biorthogonal}. In the non-interacting limit, the spectrum of $H$ in the presence of NHSE is known to be drastically different under different boundary conditions: in the thermodynamic limit, the spectrum of $H$ is purely real under open boundary condition (OBC), and features a point gap under periodic boundary condition (PBC). On the contrary, the spectrum of the Hermitrized Hamiltonian $\mathcal{H}$ is not sensitive to boundary conditions: in the thermodynamic limit, the spectrum is the same under both boundary conditions except that when $\omega$ lies in the point gap of $H$, the spectrum under OBC shows more topological zero modes than that under PBC\cite{okuma2020topological,PhysRevA.99.052118}. It is due to the contribution of these additional zero modes to the Green's function of $\mathcal{H}$ in Eq. \eqref{eq7} that allows faithfully computing the spectral function of $H$ using NHKPM\footnote{We demonstrate this with the Hatano-Nelson model\cite{PhysRevLett.77.570} in the Supplemental Material.}, despite the different sensitivities of the spectral functions of $H$ and $\mathcal{H}$ to the boundary condition. In the interacting case, the spectrum of $H$ has no point gaps in the thermodynamic limit, and the above analysis no longer holds. Thus, it is unclear whether the algorithm allows for comparable accuracy, which we examine below. It is worth noting that the case $\gamma<J$ reduces to $\gamma=0$ under a similarity transformation to $H$:
\beq
\begin{aligned}
T=e^{\sum_l l\alpha S_l^z}, \alpha=\ln r, r=\sqrt{\frac{J+\gamma}{J-\gamma}} \\
TH(J,\gamma,J_z,h_z)T^{-1}=H'(J',0,J_z,h_z).
\end{aligned}
\eeq
The new uniform exchange is $J'=\sqrt{(J+\gamma)(J-\gamma)}$ which is approximately $J$ for $\gamma=0.1J$. This similarity transformation changes the spectral function Eq. \eqref{eq4} for $H$ to the same spectral function for $H'$, leading to dynamical correlators
analogous to the $\gamma=0$ case for $\gamma\ll J$. As a benchmark, we compute both $\tilde{\mathcal{S}}_\text{tot}(\omega)$ and $\tilde{\rho}(E,l)$ for $H$ with $L=8$, $h_z=2J$ and $\gamma=0.1J$ under OBC (Figs.\ref{fig3}(a) and (c)), where the results are analogous to Figs.\ref{fig1}(d) and (g), demonstrating that the NHKPM faithfully computes both spectral functions. We also show $\tilde{\mathcal{S}}_\text{tot}(\omega)$ for $H$ under the same set of parameters under PBC, where faithful results are obtained: the two end states with almost 0 real energy under OBC vanish under PBC. We also compute $\tilde{\rho}(E,l)$ for a longer chain with $L=24$, $h_z=J$ and $\gamma=0.1J$ (Fig.\ref{fig3}(d)), where analogous results to Fig.\ref{fig2}(d) are obtained. This demonstrates the capability of NHKPM to compute spectral functions and identify topological edge modes in the presence of NHSE. We note that in the presence of NHSE, bulk states become also localized at the edge. In this case, it is the definition of the local dynamical spin correlator in Eq. \eqref{eq4} that allows to distinguish edge states from bulk states, and image edge modes in real space. Another remark is that in the presence of NHSE, the condition number\cite{belsley2005regression} of $H$ increases exponentially. Therefore, numerically determining the exact spectrum of a Hamiltonian with NHSE is often very hard and time-consuming. In our algorithm, by using Eq.\eqref{eq7}, this hard task is converted to the computation of the Green's function of a Hermitian Hamiltonian, which is relatively more controllable.

We now comment on several future perspectives for the NHKPM algorithm. It is worth noting that in some NH systems the physically relevant state could be the state with the largest imaginary energy, which is the steady state of the system after a long time evolution. The algorithm could identify the energy of many-body excitations near that state, and applies efficiently as long as the state is not highly-entangled. In addition to addressing topology, the NHKPM algorithm can also address (higher-order) NHSE\cite{PhysRevB.102.205118} by computing spectral functions defined in a basis of solely left (right) eigenvectors, or by computing the Green's function\cite{PhysRevB.103.L241408}. Finally, we note while our algorithm is implemented within
the MPS formalism, this methodology can be extended to more generic tensor-network algorithms\cite{RevModPhys.93.045003,Orus2014} and neural-network quantum states\cite{Carleo2017,PhysRevB.104.205130}. Such extensions would allow tackling many-body interacting two-dimensional non-Hermitian problems.

To summarize, we developed a NHKPM algorithm that can efficiently compute local spectral functions in a NH many-body model. In particular, we showed that this methodology
allows to compute local dynamical spin correlators in a non-Hermitian interacting system featuring
topological excitations, directly
allowing to image the topological end modes in real and frequency space.
We further showed 
that this algorithm works in the presence of the non-Hermitian skin effect. Our results put forward a new method to efficiently compute spectral functions of non-Hermitian many-body systems of large sizes beyond the capability of exact diagonalization, and make an essential step toward the critical open problem of addressing non-trivial topology in non-Hermitian many-body systems.

\textbf{Acknowledgements}
We acknowledge
the computational resources provided by
the Aalto Science-IT project,
and the
financial support from the
Academy of Finland Projects No.
331342 and No. 336243
and the Jane and Aatos Erkko Foundation.
We thank Z. Wang, S. Sayyad, T. Hyart, M. Niedermeier,
and P. Vecsei for fruitful discussions. The numerical calculations are done with the ITensor library\cite{itensor}.

\begin{appendix}

\end{appendix}

\end{document}


\author{Guangze Chen}
\affiliation{Department of Applied Physics, Aalto University, 02150 Espoo, Finland}

\author{Fei Song}
\email{songf18@mails.tsinghua.edu.cn}
\affiliation{Institute for Advanced Study, Tsinghua University, Beijing, 100084, China}

\author{Jose L. Lado}
\email{jose.lado@aalto.fi}
\affiliation{Department of Applied Physics, Aalto University, 02150 Espoo, Finland}

\title{Supplemental Material for ``Topological spin excitations in non-Hermitian spin chains with a generalized kernel polynomial algorithm"}

\date{\today}

\maketitle

\section{Details of the NHKPM method}
We derived in the main text that the spectral function
\beq\label{eq1}
f(\omega)=\langle \psi_\text{L}|\delta^2(\omega-H)|\psi_\text{R}\rangle
\eeq
can be computed with
\beq \label{eqS2}
f(\omega)=\frac{1}{\pi}\partial_{\omega^*}G(E=0)
\eeq
where
\beq
G(E)=\langle\text{L}|(E-\mathcal{H})^{-1}|\text{R}\rangle
\eeq
is an entry of the Green's function of the Hermitrized Hamiltonian $\mathcal{H}$ (Eq. (6) in the main text) with
\beq
|\text{L}\rangle=\left(\begin{array}{c}0\\|\psi_\text{L}\rangle\end{array}\right),|\text{R}\rangle=\left(\begin{array}{c}|\psi_\text{R}\rangle\\0\end{array}\right).
\eeq
Since $G(E)$ is a function of a single variable, we can apply KPM to compute $G(E)$, let
\beq \label{eqS5}
G(E)=B(E)-iA(E),
\eeq
where $A(E)=\langle\text{L}|\delta(E-\mathcal{H})|\text{R}\rangle$ and $B(E)$ is its Hilbert transform:
\beq \label{eqS6}
B(E)=\frac{1}{\pi}\mathcal{P}\int_{-\infty}^\infty \text{d}E'\frac{A(E')}{E-E'},
\eeq
where $\mathcal{P}$ denotes the Cauchy principal value. Performing a scaling on $\mathcal{H}$: $\mathcal{H}\to\mathcal{H}/\Delta$ such that its spectrum lies in $(-1,1)$, we can perform a Chebyshev expansion on $A(E)$\footnote{Although $A(E)$ is a complex-valued function, we can do Chebyshev expansion for its real and imaginary parts, respectively, then adding up the coefficients to get Eq. \eqref{eqS7} }:
\beq \label{eqS7}
A(E)=\frac{1}{\pi\sqrt{1-E^2}}\left(\mu_0+2\sum_{n=1}^\infty\mu_nT_n(E)\right),
\eeq
where 
\beq \label{eqS8}
\mu_n=\int_{-1}^1\text{d}EA(E)T_n(E)=\langle\text{L}|T_n(\mathcal{H})|\text{R}\rangle,
\eeq
and $T_n(x)=\cos(n\arccos{x})$ satisfying the following recursion relation:
\beq \label{eqS9}
\begin{aligned}
T_0(x)=1, T_1(x)=x,
\\ T_{n+1}(x)=2xT_n(x)-T_{n-1}(x).
\end{aligned}
\eeq
Using
\beq
\mathcal{P}\int_{-1}^1\frac{T_n(y)\text{d}y}{(y-x)\sqrt{1-y^2}}=\pi U_{n-1}(x),
\eeq
together with Eqs. \eqref{eqS6} and \eqref{eqS7}, we have
\beq \label{eqS11}
B(E)=\frac{2}{\pi}\sum_{n=1}^\infty\mu_nU_{n-1}(E)
\eeq
where $U_n(x)=\sin[(n+1)\arccos{x}]/\sin[\arccos{x}]$. Combining Eqs. \eqref{eqS5},\eqref{eqS7} and \eqref{eqS11}, we have
\beq \label{eqS12}
G(E=0)=\frac{2}{\pi}\sum_{n=1}^\infty(-1)^{n+1}\mu_{2n-1},
\eeq
which can be computed by noting that $\mu_n=0$ for even $n$ and $T_n(0)=U_n(0)=0$ for odd $n$. Combining Eq. \eqref{eqS2}, Eq. \eqref{eqS8} and Eq. \eqref{eqS12}, we have Eq. (8) in the main text:
\beq\label{eqS13}
f(\omega)=\frac{2}{\pi^2}\sum_{n=1}^\infty(-1)^{n+1}\langle\text{L}|\partial_{\omega^*}T_{2n-1}(\mathcal{H})|\text{R}\rangle,
\eeq
with the recursion relation
\beq
\begin{aligned}
\partial_{\omega^*}T_{n+1}(\mathcal{H})=&2\left(\begin{array}{cc}&0\\1&\end{array}\right)T_n(\mathcal{H})\\&+2\mathcal{H}\partial_{\omega^*}T_{n}(\mathcal{H})-\partial_{\omega^*}T_{n-1}(\mathcal{H}).
\end{aligned}
\eeq
We now elaborate on the numerical details in the computation of $f(\omega)$. Let 
\beq
|A_n\rangle=T_{n}(\mathcal{H})|\text{R}\rangle
\eeq
and 
\beq
|\Psi_n\rangle=\partial_{\omega^*}T_{n}(\mathcal{H})|\text{R}\rangle.
\eeq
We can verify that
\beq
\begin{aligned}
|A_0\rangle=\left(\begin{array}{c}|\psi_\text{R}\rangle\\0\end{array}\right)&, |A_1\rangle=\left(\begin{array}{c}0\\(\omega^*-H^\dag)|\psi_\text{R}\rangle\end{array}\right)\\
|\Psi_0\rangle=\left(\begin{array}{c}0\\0\end{array}\right)&, |\Psi_1\rangle=\left(\begin{array}{c}0\\|\psi_\text{R}\rangle\end{array}\right)
\end{aligned}
\eeq
with recursion relation
\beq
\begin{aligned}
|A_{n+1}\rangle&=2\mathcal{H}|A_n\rangle-|A_{n-1}\rangle\\
|\Psi_{n+1}\rangle&=2\left(\begin{array}{cc}&0\\1&\end{array}\right)|A_n\rangle+2\mathcal{H}|\Psi_n\rangle-|\Psi_{n-1}\rangle.
\end{aligned}
\eeq
Let
\beq
\begin{aligned}
|A_{2n}\rangle=\left(\begin{array}{c}|\alpha_{2n}\rangle\\0\end{array}\right)&, |A_{2n+1}\rangle=\left(\begin{array}{c}0\\|\alpha_{2n+1}\rangle\end{array}\right)\\
|\Psi_{2n}\rangle=\left(\begin{array}{c}|\psi_{2n}\rangle\\0\end{array}\right)&, |\Psi_{2n+1}\rangle=\left(\begin{array}{c}0\\|\psi_{2n+1}\rangle\end{array}\right)
\end{aligned}
\eeq
where $n\in\mathbb{Z}$ and $n\geq0$. We thus arrive at the following recursion relation
\beq \label{eqS20}
\begin{aligned}
|\alpha_{2n}\rangle&=2(\omega-H)|\alpha_{2n-1}\rangle-|\alpha_{2n-2}\rangle\\
|\alpha_{2n+1}\rangle&=2(\omega^*-H^\dag)|\alpha_{2n}\rangle-|\alpha_{2n-1}\rangle\\
|\psi_{2n}\rangle&=2(\omega-H)|\psi_{2n-1}\rangle-|\psi_{2n-2}\rangle\\
|\psi_{2n+1}\rangle&=2|\alpha_{2n}\rangle+2(\omega^*-H^\dag)|\psi_{2n}\rangle-|\psi_{2n-1}\rangle
\end{aligned}
\eeq
with
\beq
\begin{aligned}
|\alpha_0\rangle=|\psi_\text{R}\rangle&, |\alpha_1\rangle=(\omega^*-H^\dag)|\psi_\text{R}\rangle\\ 
|\psi_0\rangle=0&, |\psi_1\rangle=|\psi_\text{R}\rangle,
\end{aligned}
\eeq
and Eq.\eqref{eqS13} becomes
\beq\label{eq12}
f(\omega)=\frac{2}{\pi^2}\sum_{n=1}^\infty(-1)^{n+1}\langle \psi_L|\psi_{2n-1}\rangle.
\eeq
In practice, we do a truncation of Eq. \eqref{eq12} to order $N$:
\beq\label{eqS23}
f_\text{KPM}(\omega)=\frac{2}{\pi^2}\sum_{n=1}^N(-1)^{n+1}g^{2N}_{2n-1}\langle \psi_L|\psi_{2n-1}\rangle.
\eeq
where 
\beq
\begin{aligned}
g^{2N}_n=\frac{(2N-n+1)\cos\frac{n\pi}{2N+1}+\sin\frac{n\pi}{2N+1}\cot\frac{\pi}{2N+1}}{2N+1}\quad\quad
\end{aligned}
\eeq
is the Jackson kernel to suppress Gibbs oscillations and improve the accuracy\cite{RevModPhys.78.275}. We note that Eq. \eqref{eqS23} can also be written as:
\beq
f_\text{KPM}(\omega)=\frac{1}{\pi}\partial_{\omega^*}G_\text{KPM}(E=0)
\eeq
where
\beq \label{eqS26}
G_\text{KPM}(E=0)=\frac{2}{\pi}\sum_{n=1}^N(-1)^{n+1}g_{2n-1}\mu_{2n-1}
\eeq
is a finite-series approximation to Eq. \eqref{eqS12} with the Jackson kernel $g_n$. The truncation to the $N^\text{th}$ order polynomial with a Jackson kernel is known to provide a Gaussian approximation to the Dirac delta function\cite{RevModPhys.78.275}:
\beq \label{eqS27}
\left(\delta(x)\right)_\text{Jackson}\approx\frac{1}{\sqrt{2\pi\sigma^2}}e^{-\frac{x^2}{2\sigma^2}},
\eeq
where $\sigma=\pi/N$. Performing a Hilbert transform on both sides of Eq. \eqref{eqS27}, we have:
\beq \label{eqS28}
\left(\frac{1}{x}\right)_\text{Jackson}\approx\frac{2}{\sqrt{2\sigma^2}}F\left(\frac{x}{\sqrt{2\sigma^2}}\right)
\eeq
where $F(x)=\exp(-x^2)\int_0^x\exp(t^2)dt$ is the Dawson function\cite{https://doi.org/10.1112/plms/s1-29.1.519}. Eq. \eqref{eqS28} provides a good approximation for $1/x$ for $x\gtrsim2\sigma$. Now, for $G(E=0)$, the KPM procedure provides an approximation:
\beq \label{eqS29}
\begin{aligned}
G(E=0)&=-\langle\text{L}|\mathcal{H}^{-1}|\text{R}\rangle\\
&=-\sum_n\langle\text{L}|\varphi_n\rangle\mathcal{E}_n^{-1}\langle\varphi_n|\text{R}\rangle\\
&\approx-\sum_n\langle\text{L}|\varphi_n\rangle\left(\mathcal{E}_n^{-1}\right)_\text{Jackson}\langle\varphi_n|\text{R}\rangle,
\end{aligned}
\eeq
where $\mathcal{E}_n$ and $|\varphi_n\rangle$ are the eigen-decomposition of $\mathcal{H}$. Eq. \eqref{eqS29} is a good approximation as long as
\beq \label{eqS30}
\sigma\lesssim\frac{\min(\mathcal{E}_n)}{2\Delta},
\eeq
where $\Delta$ is the factor we divide $\mathcal{H}$ with to keep its spectrum in $(-1,1)$. A sufficiently small $\sigma$ can be achieved by increasing $N$ in our computation.

Due to the derivative in Eq. \eqref{eqS2}, a quantitative analysis of the approximation we did for $f(\omega)$ is difficult. Qualitatively, delta functions in $f(\omega)$ are smeared in $f_\text{KPM}(\omega)$ with a width $d\approx\frac{\pi\Delta}{N}$, where $\Delta$ is the scaling factor to $\mathcal{H}$ to make its spectrum in $(-1,1)$. For $L=8$, $\Delta=6$ is sufficient in our computation, and for $L=24$, we take $\Delta=16$.

\begin{figure}[t!]

\center
\includegraphics[width=\linewidth]{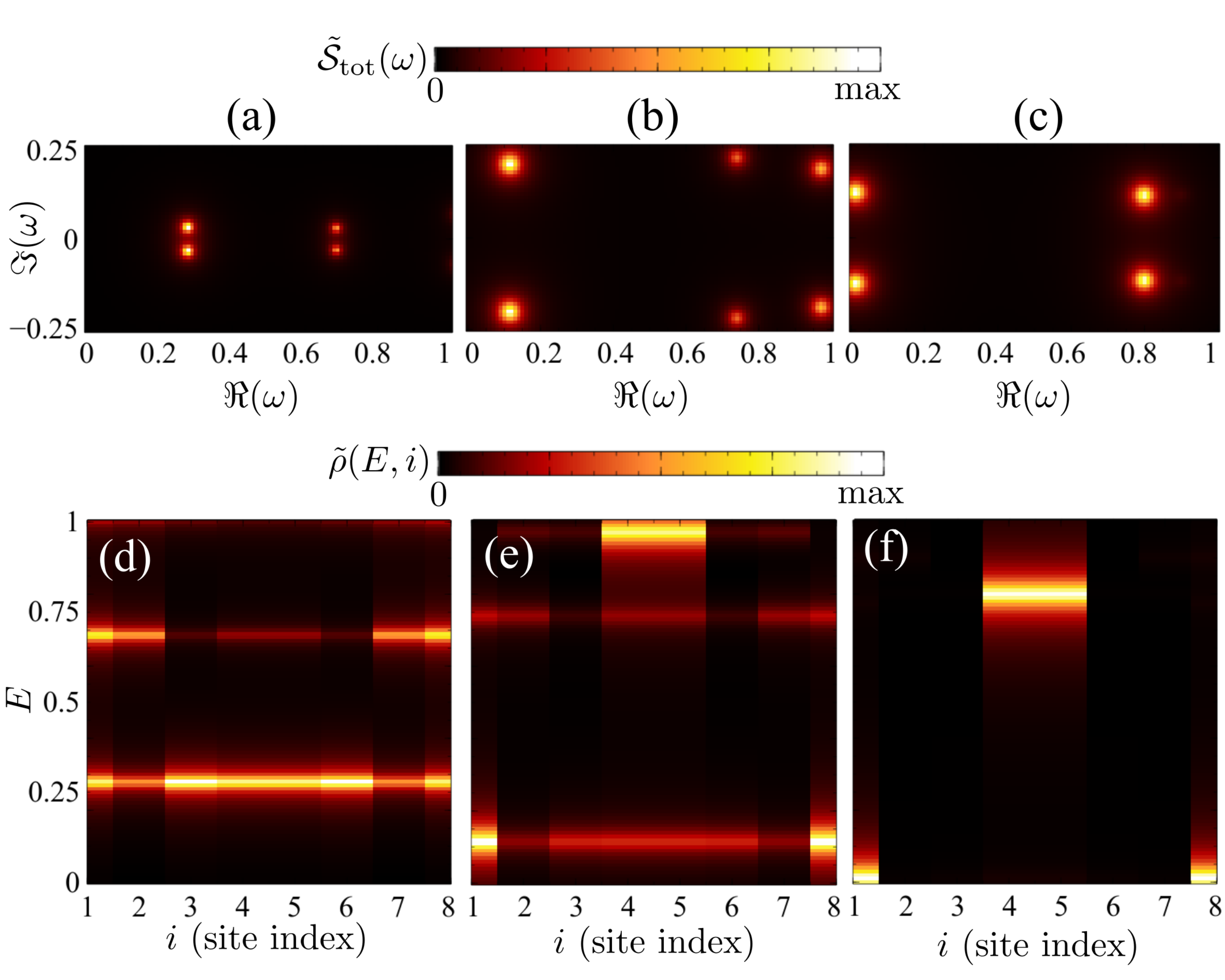}
\caption{Spectral functions in Fig.1 in the main text, computed with exact diagonalization.
}
\label{figS1}
\end{figure}

For $L=8$, the low-energy spectra computed with NHKPM agree with results obtained with exact diagonalization (ED) qualitatively (Fig.\ref{figS1}). An exact correspondence does not exist here since in ED the delta function is approximated with a Gaussian, whereas with NHKPM it is approximated with another peaked function as discussed above.

\section{Total density of states in Hatano-Nelson model with NHKPM}

\begin{figure}[t!]

\center
\includegraphics[width=\linewidth]{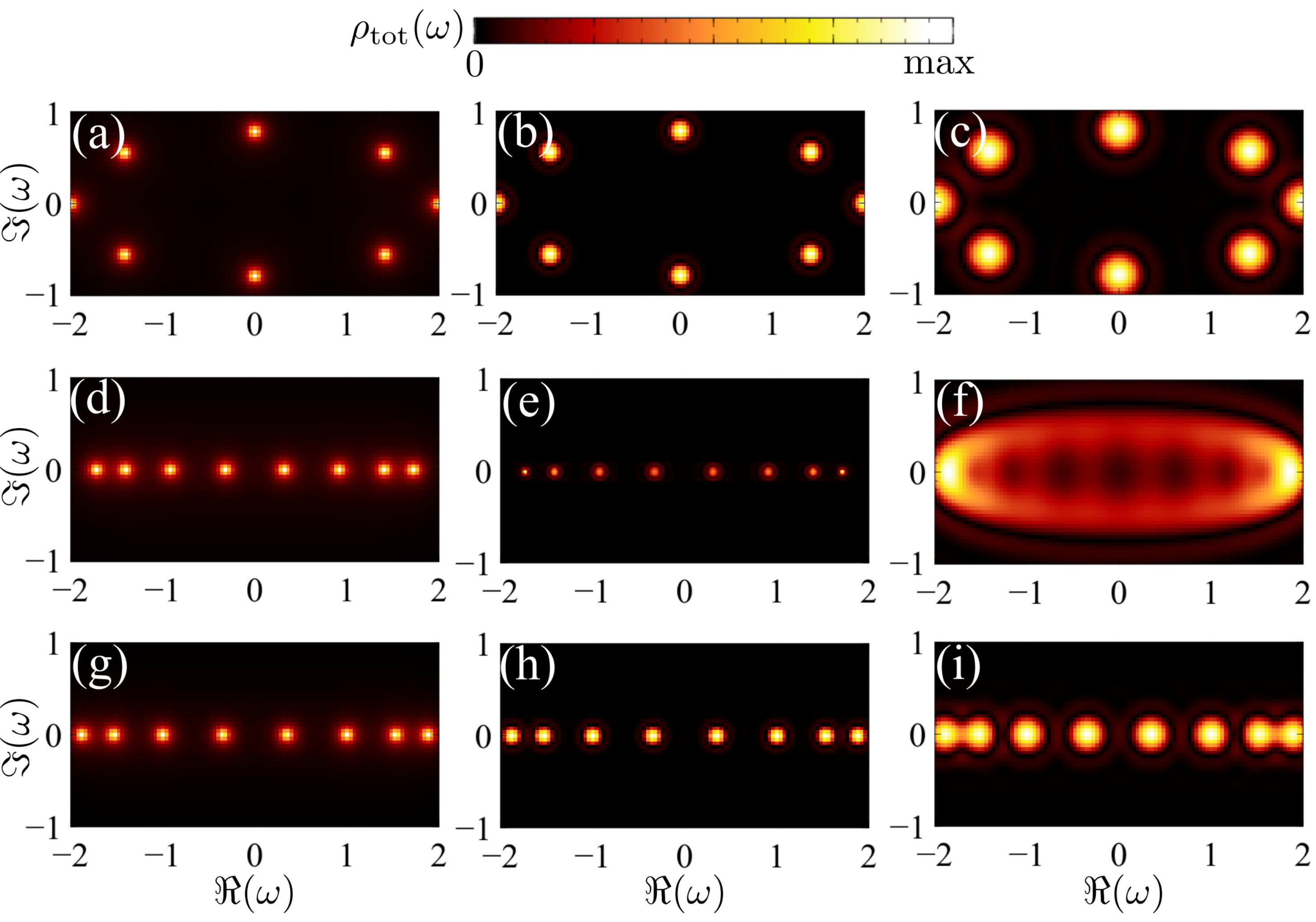}
\caption{Total density of states $\rho_\text{tot}(\omega)$ defined in Eq. \eqref{eqS32} for $H_\text{HN}$ with $L=8$ and $t=1$, with (a) $\gamma=0.4$, under PBC, computed with ED, (b) $\gamma=0.4$, under PBC, computed with NHKPM with $N=100$, (c) $\gamma=0.4$, under PBC, computed with NHKPM with $N=50$, (d) $\gamma=0.4$, under OBC, computed with ED, (e) $\gamma=0.4$, under OBC, computed with NHKPM with $N=1000$, (f) $\gamma=0.4$, under OBC, computed with NHKPM with $N=50$, (g) $\gamma=0$, under OBC, computed with ED, (h) $\gamma=0$, under OBC, computed with NHKPM with $N=100$, and (i) $\gamma=0$, under OBC, computed with NHKPM with $N=50$.
}
\label{figS2}
\end{figure}

To show that the NHKPM faithfully computes spectral functions in the single particle case in the presence of non-Hermitian skin effect (NHSE), we consider the Hatano-Nelson model\cite{PhysRevLett.77.570}:
\beq\label{eqS31}
H_{\text{HN}}=\sum_{l=1}^{L-1}(t+\gamma)c^\dag_lc_{l+1}+(t-\gamma)c^\dag_{l+1}c_{l},
\eeq
which exhibits NHSE when $\gamma\neq0$. We compute the total density of states of the model:
\beq\label{eqS32}
\rho_{\text{tot}}(\omega)=\sum_n\delta(\Re(\omega-E_n))\delta(\Im(\omega-E_n)).
\eeq
For simplicity we take $L=8$, and we note that analogous results can be obtained for larger $L$. The spectrum of $H_{\text{HN}}$ computed with ED with $t=1$ and $\gamma=0.4$ under periodic boundary condition (PBC) and open boundary condition (OBC) are shown in Figs.\ref{figS2}(a) and (d). We see that the PBC spectrum features a point gap, whereas the OBC spectrum is purely real. The spectrum computed with NHKPM in both cases are shown in Figs.\ref{figS2}(b) and (e), demonstrating the capability of NHKPM to provide faithful results in both cases. As discussed in the main text, the reason that NHKPM provides a faithful OBC spectrum in this case is due to the contribution of the zero modes of $\mathcal{H}$ to $G(E=0)$ when $\omega$ lies in the point gap of $H$. Due to finite-size effects, these zero modes have an exponentially small finite energy: $\mathcal{E}_0\propto \exp(-L/L_0)$ where $L_0$ is a constant. According to Eq. \eqref{eqS30}, to accurately capture the contribution of this zero mode to $G(E=0)$, we require:
\beq \label{eqS33}
\sigma=\frac{\pi}{N}\lesssim\frac{\mathcal{E}_0}{2\Delta},
\eeq
which for the exponentially small $\mathcal{E}_0$ requires an exponentially large $N$, hence the large $N$ in Figs.\ref{figS2}(e). When smaller $N$ is used, the computed spectrum becomes a pseudo-spectrum\cite{PhysRevB.102.014203} under OBC (Fig.\ref{figS2}(f)) due to inaccurately accounting for the contribution of the zero mode to $G(E=0)$, while the spectrum under PBC only gets broadened (Fig.\ref{figS2}(c)). For comparison we also show the OBC spectrum computed with $\gamma=0$ (Fig.\ref{figS2}(g-i)), where only a broadening is observed for smaller $N$ due to the absence of NHSE.


\section{Entanglement entropy growth during the recursive calculations}

\begin{figure}[t!]

\center
\includegraphics[width=\linewidth]{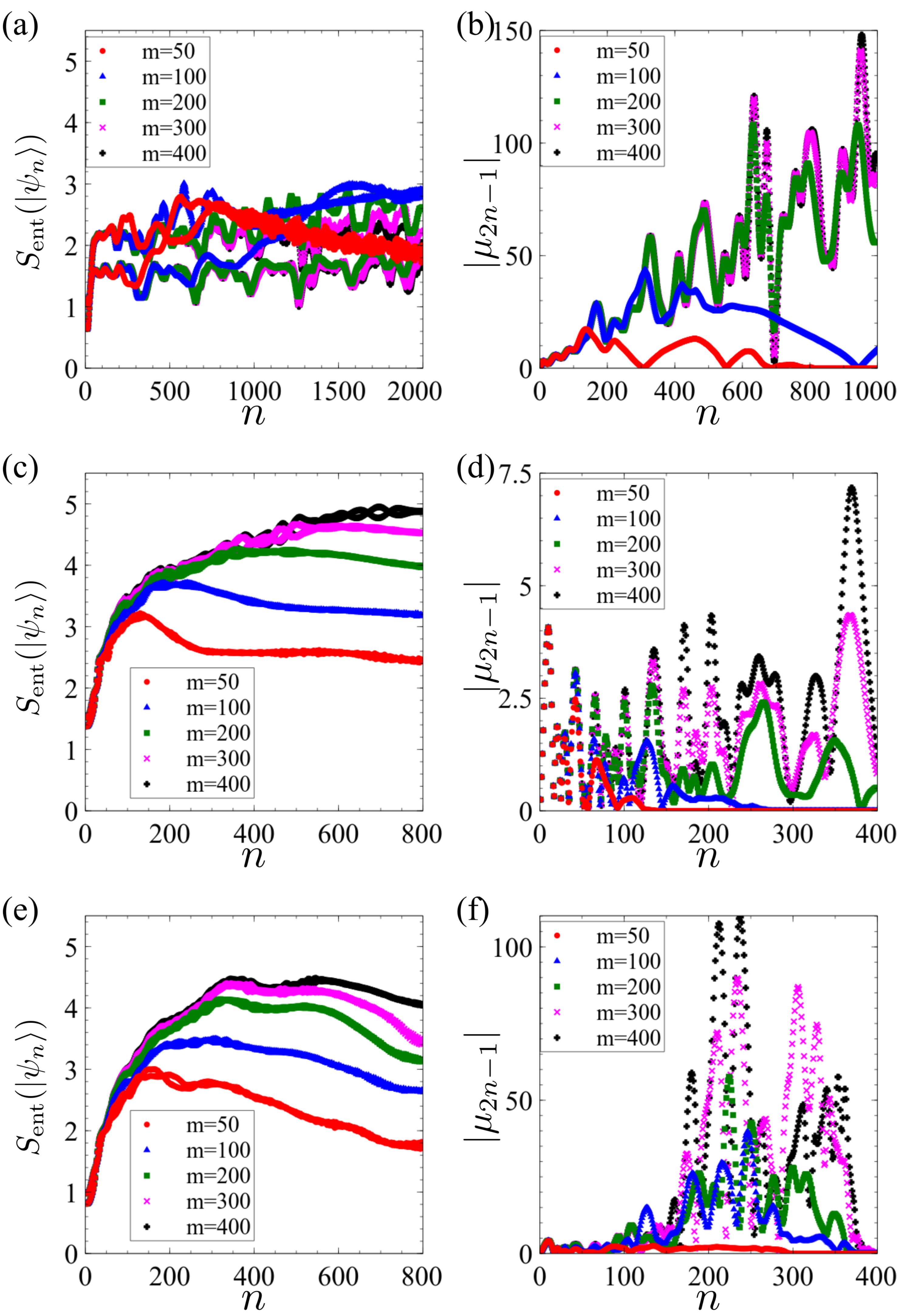}
\caption{(a) Entanglement entropy of $|\psi_n\rangle$ and (b) the coefficients $|\mu_{2n-1}|$ in the iterative calculation Eq. \eqref{eqS20} for $H$ in Eq. (9) in main text with $h_z=\gamma=0$ and $\omega=0.1$. (c,d) Same as (a,b), but with $h_z=J$, $\gamma=0$ and $\omega=0.23i$. (e,f) Same as (a,b), but with $h_z=J$, $\gamma=0.1J$ and $\omega=0.23i$.
}
\label{figS3}
\end{figure}

The updating of the states $|\psi_n\rangle$ in Eq. \eqref{eqS20} can result in states with large entanglement for large $n$ even if the initial state $|\psi_\text{R}\rangle$ is not much entangled. A largely entangled state would require larger bond dimensions in the matrix-product state (MPS) representation of the states, which increases computational cost. We analyze this entanglement increase in our calculation of the spectral function Eq.(4) for the Hamiltonian Eq.(9) in the main text, and we focus on the case $L=24$. For concreteness we choose $|\psi_\text{L,R}\rangle=S_l^x|\text{GS}_\text{L,R}\rangle$ with $l=12$, and the choice of $\omega$ is different in each specific case. We note that the results are qualitatively independent of the choice of site index $l$ and $\omega$. 

The entanglement of a certain state $|\psi\rangle$ is characterized by the entanglement entropy at the middle of the chain $S_\text{ent}$, defined as
\beq \label{eqS34}
S_\text{ent}(|\psi\rangle)=\text{Tr}(\rho_\text{A}\log\rho_\text{A})\\
\label{eqS35}
\rho_\text{A}=\text{Tr}_\text{B}(|\psi\rangle\langle\psi|),
\eeq
where A and B are the left half and right half of the chain. The entanglement entropy defined in this way is meant to measure the numerical complexity to capture the wavefunction. In particular, for an MPS, it provides a lower bound for the bond dimension $m$\cite{SCHOLLWOCK201196}:
\beq
S_\text{ent}\leq S_\text{max}=-\sum_{i=1}^m \frac{1}{m}\log_2(\frac{1}{m})=\log_2m,
\eeq
where $S_\text{max}$ is the largest possible entanglement entropy of a bond of dimension $m$. Thus, a larger $S_\text{ent}$ indicates that a larger bond dimension is needed to accurately represent the state with MPS.

We first analyze $S_\text{ent}$ for $|\psi_n\rangle$ in the Hermitian case, and with different bond dimensions $m$ used in the MPS representation (Fig.\ref{figS3}(a)). It is first observed that for $m=50$ and $m=100$ the entanglement entropy starts to differ from results computed with a higher bond dimension at around $n=500$, indicating that $m$ is too small to accurately represent the state. Another observation is that the entanglement entropy saturates to a small value around 2 for large $n$, allowing the states to be represented accurately using a small $m$. As a consequence, the coefficients $\mu_{2n-1}$ computed with $m=200$ is accurate up to $n=600$, and with $m=300$ is accurate up to $n=1000$ (Fig.\ref{figS3}(b)). When a large non-Hermitian term $h_z=J$ is considered, the entanglement increase in the iterations becomes faster, and the entanglement entropy grows to a much larger value compared to the Hermitian case (Fig.\ref{figS3}(c)). As a consequence, the number of faithful coefficients drops for the same bond dimension $m$ used (Fig.\ref{figS3}(d)). In the Hermitian case, Figs.2(a,c) in the main text are computed with $N=500$ and $m=200$. In the case $h_z=J$, Figs.2(b,d) in the main text are computed with $N=300$ and $m=300$.

The presence of the non-Hermitian skin effect also results in a faster increase of entanglement entropy of $|\psi_n\rangle$ (Fig.\ref{figS3}(e)), and the faithful number of coefficients for fixed $m$ further drops (Fig.\ref{figS3}(f)). As a consequence, Fig.3(d) in the main text is computed with $N=200$ and $m=400$.

\section{Numerical complexity of NHKPM}
We analyze the numerical complexity of our algorithm in this section. In essence, an optimized version of our algorithm scales approximately
as $L^3$, with $L$ the number of sites. This power law dependence
makes our algorithm scalable to bigger systems. We elaborate on the details below.

We analyze the time consumption to compute the spectral function (Eq.(4) in the main text):
\beq \label{eqR6}
\begin{aligned}
\mathcal{S}(\omega,l)=&\langle \text{GS}_\text{L} | S_l^{-}\delta^2(\omega+E_\text{GS}-H)S_l^{+} |\text{GS}_\text{R}\rangle\\
&+\langle \text{GS}_\text{L} | S_l^{+}\delta^2(\omega+E_\text{GS}-H)S_l^{-} |\text{GS}_\text{R}\rangle
\end{aligned}
\eeq
with NHKPM for different system size $L$, where $H$ and $|\text{GS}_{\text{L},\text{R}}\rangle$ are given. In our particular case, the Hamiltonian $H$ is given by Eq.(9) in the main text, and $|\text{GS}_{\text{L},\text{R}}\rangle$ are computed with the Krylov-Schur algorithm. We focus on the topologically non-trivial regime: $h_z=J,\gamma=0$, and fix $\omega=0.23i$ and $l=1$ as the spectral function shows a peak at this point due to the topological edge excitations. We note that the energy mesh and a sweep over all sites required to compute the full spectral function can be easily parallelized.
As a result, the scaling of the algorithm is
determined by the size dependence for a fixed $\omega$ and $l$.

We fix the broadening of the peak for the scaling analysis. As the system size increases, a larger scaling factor $\Delta$ is required to scale the spectrum of $\mathcal{H}$ into $(-1,1)$. Thus, to ensure the same broadening, the number of polynomials $N$ should correspondingly increase. For our specific model, $\Delta$ is proportional to $L$, and thus we choose $N\propto L$ for the scaling analysis.

\begin{figure}[t!]
\center
\includegraphics[width=\linewidth]{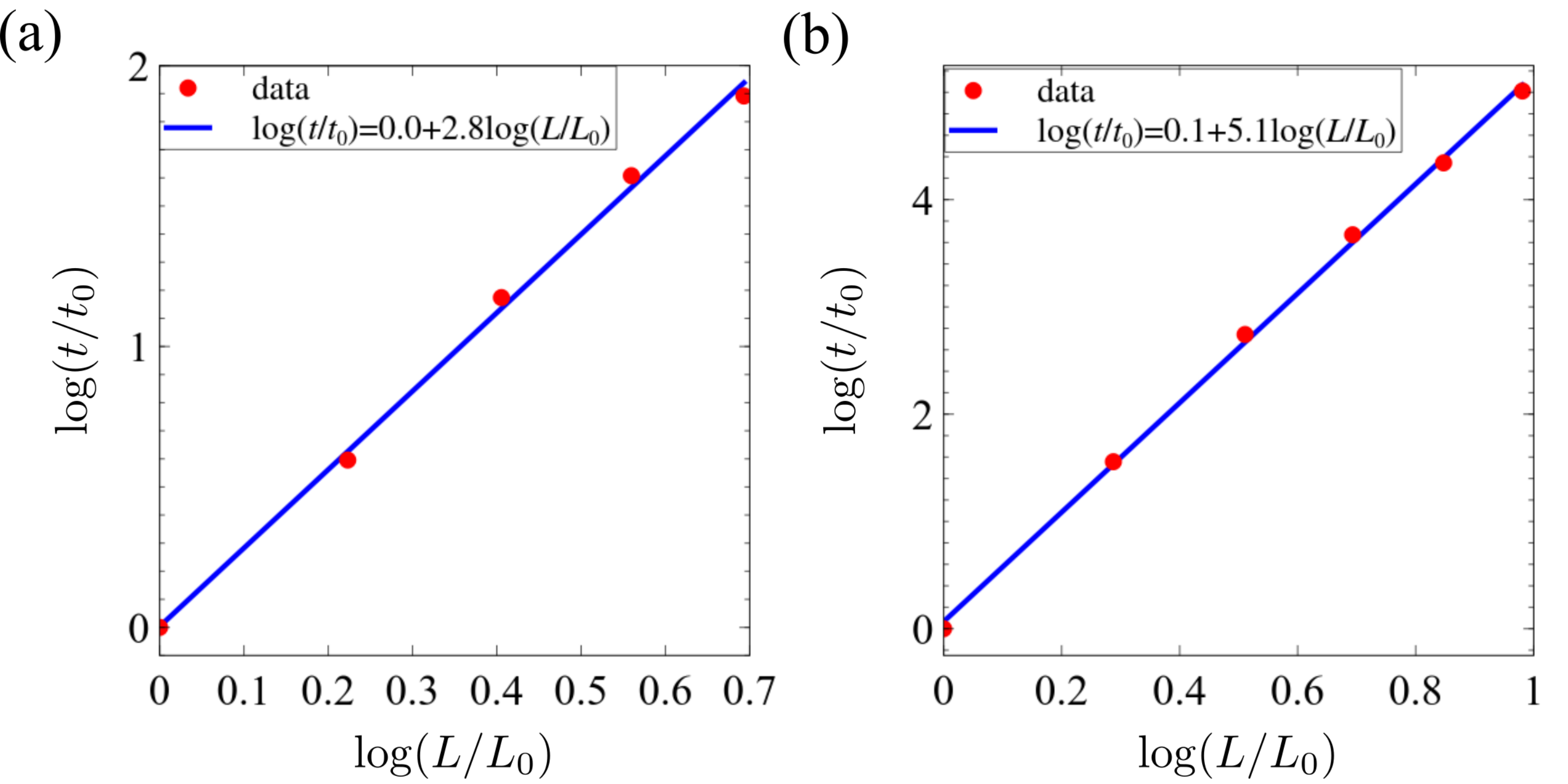}
\caption{(a) Complexity of NHKPM algorithm with fixed bond dimension $m=80$ in MPS computations and $N=25L/2$, with $L_0=16$ and $t_0=442.9$s\footnote{The data is obtained with the Julia @btimes function, which averages time consumption over several computations. The computations are done on \href{https://scicomp.aalto.fi/triton/overview/}{Aalto University's triton cluster node pe84}, with the CPU Intel Xeon E5 2680 v4.}. The log-log data is fitted with a linear function, indicating that the complexity of the algorithm is power-law with $(t/t_0)\propto (L/L_0)^{2.8}$. (b) Complexity of NHKPM algorithm with bond dimension $m=10L/3$ in MPS computations and $N=25L/2$, with $L_0=12$ and $t_0=30.4$s\footnote{The computations are done on \href{https://scicomp.aalto.fi/triton/overview/}{Aalto University's triton cluster node c604}, with the CPU Intel Xeon E5 2690 v3.}. The log-log fit indicates a power-law complexity with $(t/t_0)\propto (L/L_0)^{5}$.
}
\label{figS4}
\end{figure}

We first analyze the scaling of our algorithm for a generic case where the bond dimension of the MPS $m$ is fixed. In this case, the time consumption shows an approximately 3rd power dependence on the system size (Fig.\ref{figS4}(a)). This scaling is the same as the kernel polynomial algorithm for Hermitian interacting systems\cite{PhysRevResearch.1.033009}. 

\begin{figure}[t!]
\center
\includegraphics[width=\linewidth]{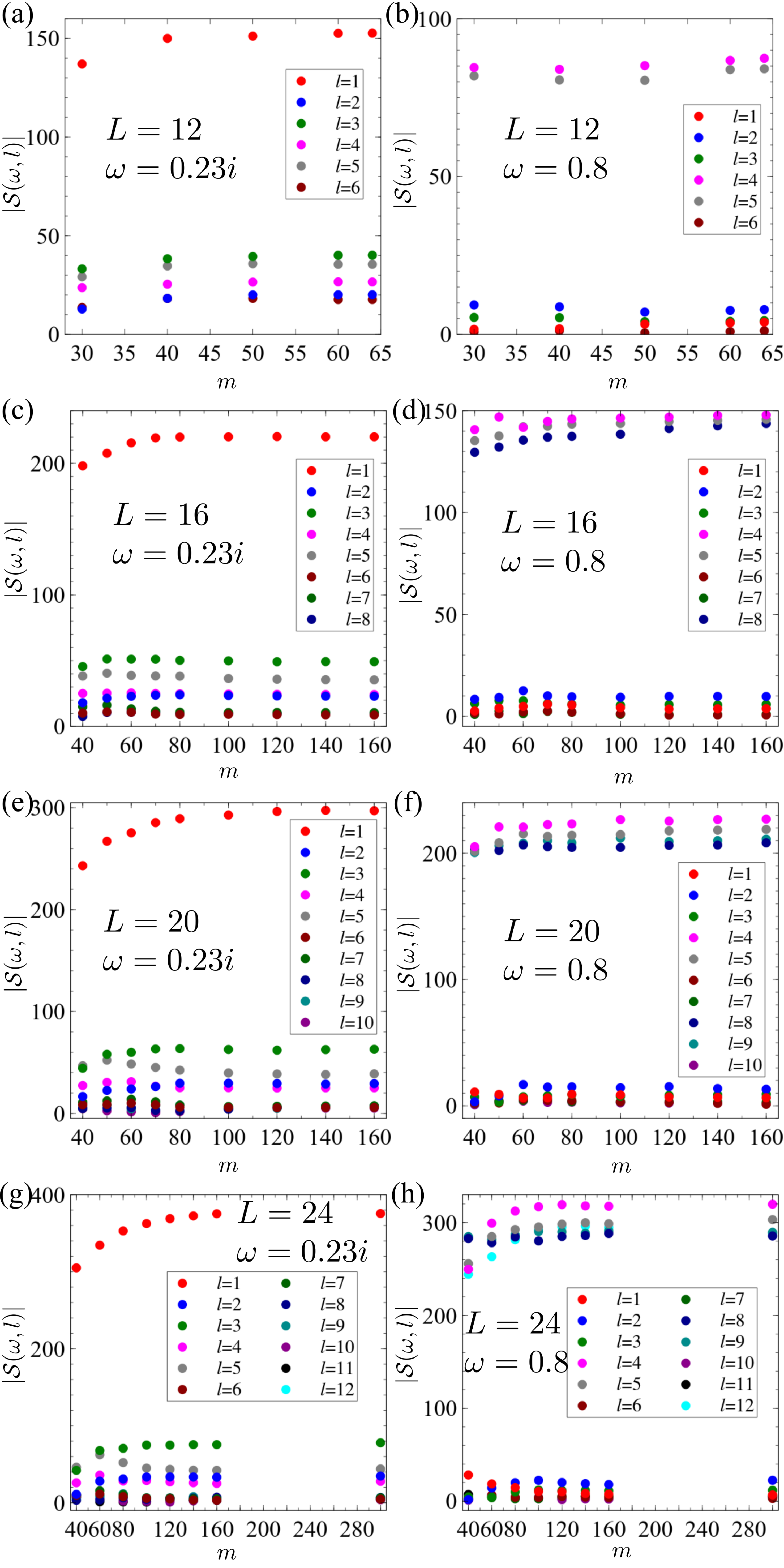}
\caption{The spectral function $\mathcal{S}(\omega,l)$ computed for different $L$ and $\omega$ as a function of maximum bond dimension $m$. The parameters for $H$ (Eq. (9) in the main text) are $h_z=J$ and $\gamma=0$.}
\label{figS5}
\end{figure}

We also analyze the scaling for our specific computation, where the entanglement entropy of states increases during the recursive calculations in the kernel polynomial method. In this case, to ensure the computational accuracy, a sufficiently large $m$ is required, which increases with $L$. In Fig.\ref{figS5}, we show the computed spectral functions as a function of $m$ for different sizes $L$. We show this for both $\omega=0.23i$ where the spectral function is dominated at the boundary, and for $\omega=0.8$ where the spectral function is dominated in the bulk. We observe that as $m$ increases, the computed spectral function approaches the accurate value as expected. We also observe that although the accuracy is low for small $m$, it is still clear that for $\omega=0.23i$ there exist edge excitations, and for $\omega=0.8$ the excitations are more distributed in the bulk. Thus, a relative error of $10\%$ is sufficient for the characterization of topological edge states. The smallest $m$ required for this accuracy is $m=40,50,60,80$ for $L=12,16,20,24$, respectively. For simplicity, we choose $m=10L/3$ to study the scaling, resulting in computational time $t\propto L^5$ (Fig.\ref{figS4}(b)). Compared to the fixed $m$ case, the additional power of 2 originates from the fact that the complexity of applying a matrix product operator (MPO) on an MPS is proportional to $m^2$, and we have used $m\propto L$ in this case. 

\section{Fidelity of ground states computed with the Krylov-Schur algorithm}

The ground states are computed with the Krylov-Schur algorithm\cite{doi:10.1137/S0895479800371529} in our case using matrix product state algebra. This method relies on an iterative algorithm to find the Krylov subspace expanded by a set of matrix-product states. Compared to generalized DMRG algorithms\cite{TEnss_2001,doi:10.1063/1.1899124,PhysRevB.101.235150,PhysRevB.105.205125}, this method does not have the limitations given by the local updates of DMRG-like sweeps, and its accuracy is solely controlled
by the bond dimension. In addition, it applies to arbitrary Hamiltonians without requiring specific symmetries, which is beneficial for studying Hamiltonians with generic ground states as in the case we studied.

We first compute the Krylov space expanded by 4 eigenvectors with the smallest real eigenvalue. We then identify the one with the smallest real eigenvalue out of the 4 as the ground state. In this way, we avoid the risk of mistaking an eigenvector with a similar eigenvalue to the ground state as the ground state. The calculated ground state is then verified by checking the deviations:
\beq
d_\text{R}(|\psi_\text{R}\rangle)=\frac{|\langle \psi_\text{R}|H^\dag H|\psi_\text{R}\rangle|-|\langle \psi_\text{R}| H|\psi_\text{R}\rangle|^2}{|\langle \psi_\text{R}|H^\dag H|\psi_\text{R}\rangle|}\\
d_\text{L}(|\psi_\text{L}\rangle)=\frac{|\langle \psi_\text{L}|H H^\dag|\psi_\text{L}\rangle|-|\langle \psi_\text{L}| H^\dag|\psi_\text{L}\rangle|^2}{|\langle \psi_\text{L}|H H^\dag|\psi_\text{L}\rangle|}.
\eeq
An exact right/left eigenvector $|\psi_\text{R/L}\rangle$ of $H$ should have $d_\text{R}(|\psi_\text{R}\rangle)=0$ and $d_\text{L}(|\psi_\text{L}\rangle)=0$. For $L=8$, $h_z=J$, $d_\text{R}(|\text{GS}_\text{R}\rangle)=1.4e{-5}$ and $d_\text{L}(|\text{GS}_\text{L}\rangle)=1.0e{-5}$; and when $h_z=2J$, $d_\text{R}(|\text{GS}_\text{R}\rangle)=9.7e-5$ and $d_\text{L}(|\text{GS}_\text{L}\rangle)=9.9e-5$. For $L=24$, $d_\text{R}(|\text{GS}_\text{R}\rangle)=3.0e-4$ and $d_\text{L}(|\text{GS}_\text{L}\rangle)=4.0e-4$ for $h_z=J,\gamma=0$, and  $d_\text{R}(|\text{GS}_\text{R}\rangle)=7.2e-4$ and $d_\text{L}(|\text{GS}_\text{L}\rangle)=6.9e-4$ for $h_z=J,\gamma=0.1J$. Thus we have verified the ground states used in our calculation are faithful ground states.

\bibliography{main}